\documentclass[aps,prb,twocolumn,10pt,superscriptaddress,longbibliography,showpacs,floatfix]{revtex4-2}
\usepackage[dvipsnames]{xcolor}
\usepackage{physics}
\usepackage{amssymb}
\usepackage{graphicx}
\usepackage{mathtools}
\usepackage{amsmath}
\usepackage{hyperref}
\usepackage{tabularx}
\usepackage{comment}
\usepackage{amsfonts,dsfont}
\usepackage{braket}
\usepackage{acronym}
\usepackage{multirow}
\usepackage{booktabs}
\usepackage[export]{adjustbox}
\usepackage{enumerate}
\usepackage{bbm}

\newacro{mps}[MPS]{matrix product state}
\newacro{svd}[SVD]{singular value decomposition}
\newacro{dmrg}[DMRG]{density matrix renormalization group}
\newacro{tebd}[TEBD]{time evolution block decimation}

\begin{document}
\title{Fast Time-Evolution of Matrix-Product States using the QR decomposition}
\author{Jakob Unfried}
\email{jakob.unfried@tum.de}
\affiliation{Department of Physics, TFK, Technische Universit{\"a}t M{\"u}nchen, James-Franck-Stra{\ss}e 1, D-85748 Garching, Germany}
\affiliation{Munich Center for Quantum Science and Technology (MCQST), Schellingstr. 4, 80799 M\"unchen, Germany}
\author{Johannes Hauschild}
\affiliation{Department of Physics, TFK, Technische Universit{\"a}t M{\"u}nchen, James-Franck-Stra{\ss}e 1, D-85748 Garching, Germany}
\affiliation{Munich Center for Quantum Science and Technology (MCQST), Schellingstr. 4, 80799 M\"unchen, Germany}
\author{Frank Pollmann}
\affiliation{Department of Physics, TFK, Technische Universit{\"a}t M{\"u}nchen, James-Franck-Stra{\ss}e 1, D-85748 Garching, Germany}
\affiliation{Munich Center for Quantum Science and Technology (MCQST), Schellingstr. 4, 80799 M\"unchen, Germany}
%
%
%
%
\begin{abstract}
    We propose and benchmark a modified \ac{tebd} algorithm that uses a truncation scheme based on the QR decomposition instead of the \ac{svd}.
    The modification reduces the scaling with the dimension of the physical Hilbert space $d$ from $d^3$ down to $d^2$.
    Moreover, the QR decomposition has a lower computational complexity than the SVD and allows for highly efficient implementations on GPU hardware.
    In a benchmark simulation of a global quench in a quantum clock model, we observe a speedup of up to three orders of magnitude comparing QR and SVD based updates on an A100 GPU.
\end{abstract}
\maketitle
%
%
\section{Introduction}\label{sec:introduction}
Numerical simulations of the dynamics of quantum many-body systems in and out of equilibrium is essential for the understanding of a wide range of physical phenomena.
Following the success of the \ac{dmrg} method~\cite{White1992, Schollwock2011Density} for efficiently finding ground states of one-dimensional (1D) quantum systems in terms of matrix product states (MPS), several related techniques have been developed to efficiently simulate the time-evolution~\cite{White2004, Daley2004, Vidal2004, Haegeman2011, Haegeman2016, paeckel2019time}.
These methods have since allowed access to experimentally relevant observables, such as dynamical correlation functions which can be compared with data from neutron scattering and ultracold atomic gasses~\cite{Gohlke2017Dynamics, Kadow2022Hole, Jepsen2021Transverse}, and far out of equilibrium dynamics~\cite{Kollath2007Quench}, providing profound insights into long standing questions about quantum thermalization~\cite{trotzky2012probing}, many-body localization~\cite{Bardarson2012Unbounded, Titas2020Time, Doggen2021Many, Nietner2022Route} and transport properties~\cite{Prosen2009,Rakovszky2022,Bertini2021,Schulz2018Energy, Kloss2020Spin, Darkwah2022Probing}.
\par
In a series of recent works~\cite{Li2020, Feng2022, Hauru2021, Morningstar2022Simulation,Ganahl2022}, it has been demonstrated that accelerated linear algebra operations on  graphics processing units (GPUs) and tensor processing units (TPUs) allows various numerical tasks, and in particular simulation of quantum dynamics, to be carried out not only significantly faster but also more power efficiently.
However, many MPS based algorithms heavily rely on \acfp{svd}, which are slow in the GPU implementations known to us.
%
For example, the prominent time-evolving block decimation (TEBD)~\cite{Vidal2004,Vidal2007Classical} algorithm performs an SVD following every application of a two-site gate to truncate the bond dimension.
In this work, we propose a modification to the \ac{tebd} algorithm for \ac{mps} time evolution, which uses QR decompositions to achieve a variational truncation, replacing the \ac{svd}.
This truncation scheme is not only faster already on CPUs as it reduces the scaling with the dimension of the physical Hilbert space $d$ from $d^3$ down to $d^2$, but unlike for the \ac{svd} based scheme, significant speedups can be achieved on GPUs at the same accuracy.
\par
This paper is organized as follows:
In section~\ref{sec:algorithm}, we briefly review \ac{mps} and introduce the QR based truncation scheme.
We elaborate on a way to dynamically adjust the \ac{mps} bond dimension in section~\ref{sec:bond_dim_expansion}.
A detailed benchmark study is provided in section~\ref{sec:benchmark}, comparing results and runtimes between the different \ac{tebd} schemes, both on CPU and GPU hardware, before we conclude our findings in section~\ref{sec:conclusion}.
%
%
%
\section{QR based time evolution algorithm}\label{sec:algorithm}
\begin{figure}[t]
\vspace{0.5cm}
\includegraphics[width=0.48\textwidth]{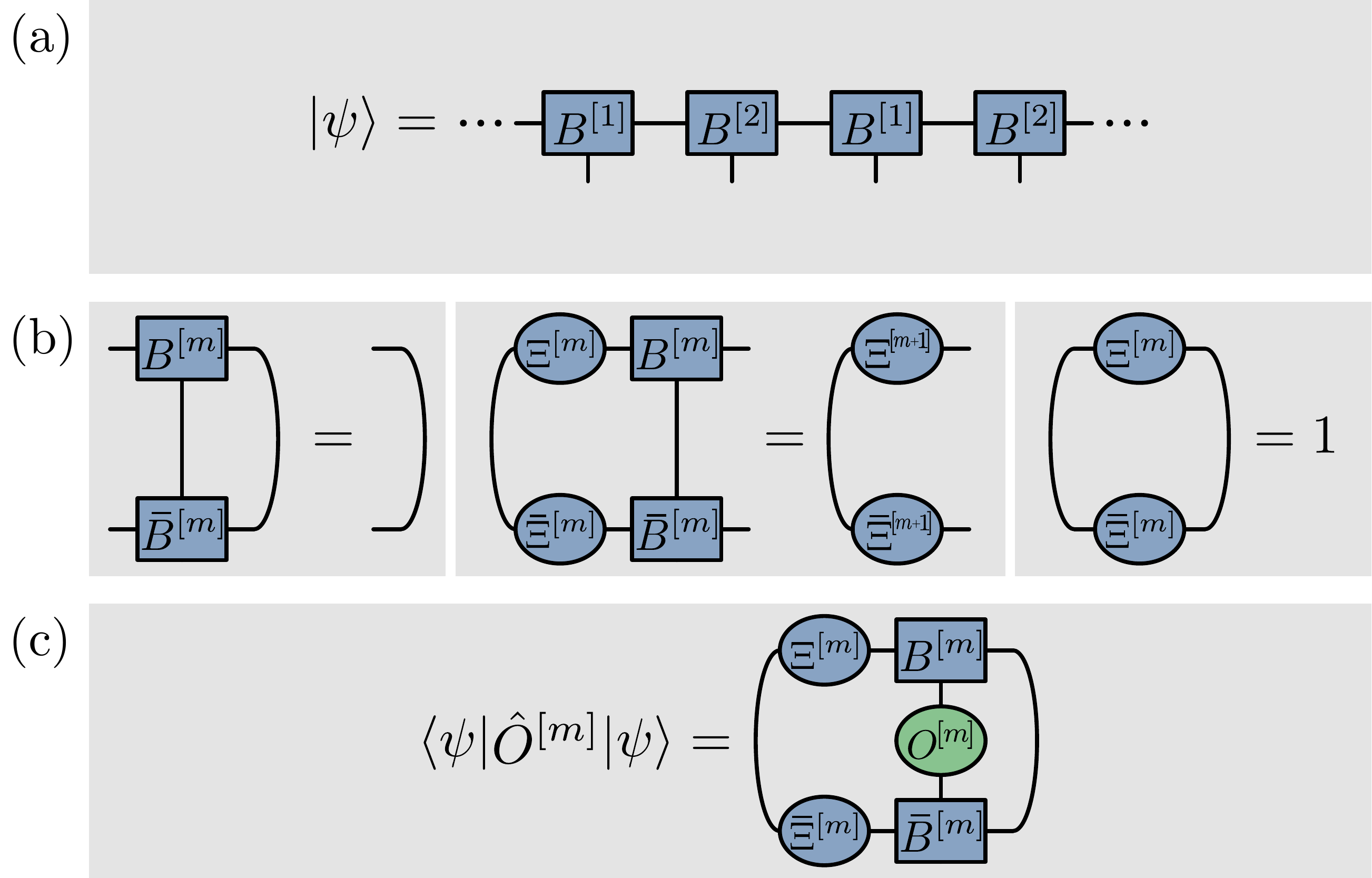}
\caption{
    \label{fig:iso_mps}
    (a) Uniform \ac{mps}, depicted here with a unit cell of two sites.
    (b) Conditions for the right isometric form; dominant right eigenvector of the transfer matrix, dominant left eigenvector, normalization choice for left eigenvectors.
    (c) The isometry conditions allow easy evaluation of local expectation values.
}
\end{figure}
We first review the isometric form of an \ac{mps} as shown schematically in Fig.~\ref{fig:iso_mps}.
While the algorithm can be used both for finite as well as for uniform (i.e., infinite) \ac{mps}, in the following we only focus on the latter case and assume a unit cell of $L$ sites.
The \ac{mps} is parametrized by matrices $B^{[m]i}$, where $i$ labels a basis of the local Hilbert space on site $m$ and matrix indices are suppressed, such that
\begin{equation}
    \ket{\psi} = \sum_{\{i_n\}} (\ldots  B^{[m]i_m} B^{[m+1]i_{m+1}} \ldots) \ket{\{i_n\}}
    .
\end{equation}
The transfer matrix for a unit cell starting on site $m$ is given by $T_m = T^{[m]}T^{[m+1]}\dots T^{[m+L-1]}$, where $T^{[m]}_{(\alpha\alpha')(\beta\beta')} = \sum_{i}B^{[m]i}_{\alpha\beta}\bar{B}^{[m]i}_{\alpha'\beta'}$.
We choose a right isometric form, in which the dominant right eigenvector $\rho^{[m]}$ of $T_m$ is $\rho^{[m]}_{\beta\beta'} = \delta_{\beta\beta'}$, while its dominant left eigenvector is given by $\lambda^{[m]}_{\alpha\alpha'} = \sum_{\beta} \Xi^{[m]}_{\beta\alpha}\bar{\Xi}^{[m]}_{\beta\alpha'}$ with $\|\Xi^{[m]}\|=1$ and both respective eigenvalues equal to one.
Moreover, $T^{[m]}$ translate the eigenvectors, i.e. $\lambda^{[m]} T^{[m]} = \lambda^{[m + 1]}$ and $T^{[m]} \rho^{[m]} = \rho^{[m - 1]}$, with superscripts modulo $L$.
Note that this makes the $B^{[m]}$ isometric, in the sense that $\sum_{i\beta} B^{[m]i}_{\alpha\beta} \bar{B}^{[m]i}_{\alpha'\beta} = \delta_{\alpha\alpha'}$.
An \ac{mps} in this form allows us to directly evaluate local expectation values.
Note that the isometric form does not fully fix the gauge freedom and is a weaker requirement than the canonical form of \ac{mps}~\cite{Vidal2003Entanglement, Vidal2007Classical}, which would additionally require that the $\Xi^{[m]}$ are diagonal matrices with real positive entries in descending order, i.e.~the Schmidt values $\Lambda_{\alpha}$ for a bipartition of the state by cutting the bond between sites $m$ and $n$, $|\psi\rangle = \sum_{\alpha=1}^\chi|\alpha\rangle_{\triangleleft}\Lambda_{\alpha}|\alpha\rangle_{\triangleright}$.
Here, $|\alpha\rangle_{\triangleleft (\triangleright)}$ denote orthonormal states on the sites left (right) of the given bond, i.e.~the left (right) Schmidt states.
%
In the isometric form we have $|\psi\rangle = \sum_{\alpha\beta}|\alpha\rangle_{\triangleleft}\Xi^{[n]}_{\alpha\beta}|\beta\rangle_{\triangleright}$, such that the Schmidt values can be obtained as the singular values of the non-diagonal $\Xi^{[n]}$.
\par
\begin{figure}[t]
    \vspace{0.5cm}
    \includegraphics[width=0.48\textwidth]{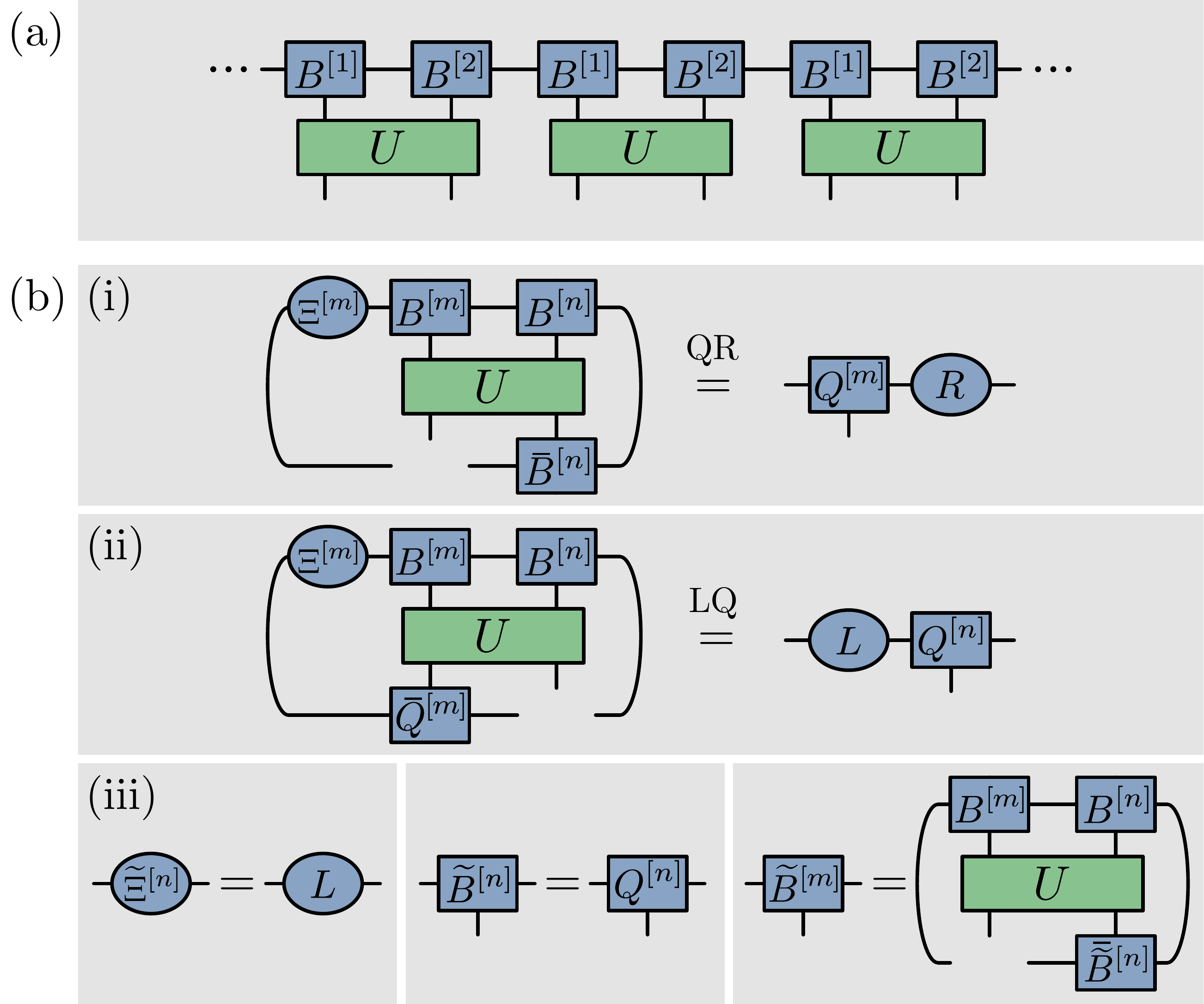}
    \caption{\label{fig:algorithm} Algorithm for the QR based time evolution:
        (a) The time evolution is decomposed into two-site gates acting on neighboring sites. (b) Algorithm for the QR based truncation scheme:
        (i) Contraction of the time evolved block with tensor $\bar{B}^{[n]}$ and subsequent QR decomposition.
        (ii)  Contraction of the time evolved block with tensor $Q^{[m]}$ obtained in the previous step and subsequent LQ decomposition.
        (iii) Obtaining the updated tensors $\tilde{B}^{[m]}$, $\tilde{B}^{[n]}$ and $\tilde{\Xi}^{[n]}$.
    }
\end{figure}

In order to approximate the time evolution of the \ac{mps} with respect to a Hamiltonian $H = \sum_n H_{n,n+1}$, we apply the Trotterized time evolution operator alternatingly to even and odd bonds as shown in Fig.~\ref{fig:algorithm}a in the same way as in the original \ac{svd} based infinite \ac{tebd} algorithm~\cite{Vidal2004,Vidal2007Classical}.
However, the update procedure for two neighboring sites $m$ and $n = m + 1 ~(\mathrm{mod}~L)$ differs in that we do not require an \ac{svd} decomposition with a cost scaling as $d^3\chi^3$ with the local Hilbert space dimension $d$ and \ac{mps} bond dimension $\chi$.
Instead, the algorithms relies on two successive QR (or LQ) decompositions and scales as $d^2\chi^3$.
As shown in Fig.~\ref{fig:algorithm}b, the algorithm consists of three steps:
\begin{enumerate}[(i)]
    \item We first construct a mixed representation $\theta_{\alpha\delta}^{ij}=\sum_{\beta,\gamma} \Xi_{\alpha\beta}B^{[m]i}_{\beta\gamma}B^{[n]j}_{\gamma\delta}$ of the state in terms of physical and virtual states.
    We apply the two-site gate $U$ to the state, $\tilde{\theta}_{\alpha\delta}^{ij} = \sum_{i'j'} U^{i'j'}_{ij}\theta_{\alpha\delta}^{i'j'}$.
    The evolved state is then projected back into the manifold of \ac{mps} of the given bond dimension, by contracting it with the complex conjugate of the isometry $B^{[n]}$ to obtain $X^{i}_{\alpha \gamma}=\sum_{j,\delta}\tilde{\theta}_{\alpha\delta}^{ij}\bar{B}^{[n]j}_{\gamma\delta}$.
    We group the legs of $X^{i}_{\alpha \gamma}\rightarrow X_{(\alpha i) \gamma}$ and perform a QR decomposition of this matrix, $X_{(\alpha i) \gamma}=\sum_{\beta} Q^{[m]}_{(\alpha i)\beta}R_{\beta\gamma}$.
    Ungrouping the legs $Q^{[m]}_{(\alpha i) \beta}\rightarrow Q^{[m]i}_{\alpha\beta}$ yields the left isometry used in the next step.

    \item We start from the evolved state $\tilde{\theta}_{\alpha\delta}^{ij}$ and project it by contracting it with the complex conjugate of the left isometry $Q^{[m]}$ to obtain $Y^{j}_{\beta \delta}=\sum_{i,\alpha}\bar{Q}^{[m]i}_{\alpha\beta}\tilde{\theta}_{\alpha\delta}^{ij}$.
    We group the legs of $Y^{j}_{\beta \delta}\rightarrow Y_{\beta (j\delta)}$ and perform an LQ decomposition of this matrix, $Y_{\beta (j\delta)}=\sum_{\gamma} L_{\beta\gamma}Q^{[n]}_{\gamma( j\delta)}$.
    Ungrouping the legs $Q^{[n]}_{\gamma (j \delta)}\rightarrow Q^{[n]j}_{\gamma\delta}$ yields the right isometry used in the next step.

    \item We conclude the iteration by assigning the updated tensors:
    $\tilde{\Xi}^{[n]}_{\beta\gamma} =  L_{\beta\gamma}$, $\tilde{B}^{[n]j}_{\gamma\delta} = Q^{[n]j}_{\gamma\delta}$,
    and $\tilde{B}^{[m]i}_{\alpha\beta} = \sum_{\gamma\delta i' j' j}U_{ij}^{i'j'}B^{[m]i'}_{\alpha\gamma}B^{[n]j'}_{\gamma\delta}\bar{\tilde{B}}^{[n]j}_{\beta\delta}$.
\end{enumerate}
A few comments are in order.
Firstly, we can understand the truncation scheme as an iterative solver for finding the optimal approximation of $\theta$ with reduced rank $\tilde\chi$, i.e.~$\tilde{\theta}_{(\alpha i)(j \delta)} \approx \sum_{\gamma=1}^{\tilde\chi} X_{(\alpha i)\gamma} Y_{\gamma(j \delta)}$.
%
Keeping one of the components, e.g.~$Y$, constant and demanding it to be an isometry, the optimal update for $X$ which minimizes the distance $\| \tilde{\theta} - XY\|$ is given by $\tilde\theta Y^\dagger$.
Before we can analogously update $Y$, we perform a gauge transformation via the QR decomposition, i.e.~$(X, Y) \mapsto (Q, RY)$, which makes the first matrix an isometry and leaves the distance invariant.
After updating $Y$, its LQ decomposition yields an approximation of $\tilde\theta$ in a suitable isometric form.
In order to approximate a generic matrix $\tilde\theta$, we expect that these updates need to be iterated until convergence.
In the specific case of \ac{mps} time evolution we have $U = \mathbbm{1} + \mathcal{O}(\delta t)$ such that the initial guess $Y_0 = B^{[n]}$ is already optimal to zeroth order in $\delta t$, and we find that a single sweep is sufficient.
\par
Secondly, the update for $\tilde{B}^{[m]}$ in the last step is motivated by Hastings' modified \ac{tebd}~\cite{Hastings2009}.
From our truncation scheme, just as from \ac{svd} based truncation, we get the left \ac{mps} tensor in left isometric form, i.e.~$\tilde A^{[m]} = Q^{[m]}$ and Hasting's modification allows us to form $\tilde{B}^{[m]i} = (\Xi^{[m]})^{-1}\tilde A^{[m]i}\tilde\Xi^{[n]}$ without explicit matrix inversion, which would in practice often be ill-conditioned.
Keeping the above left isometric update, on the other hand, allows our scheme to be applied to finite systems, where \ac{mps} tensors left of the next update are required to be in left isometric form.
\par
Lastly, the algorithm, as presented above, yields an \ac{mps} with the same bond dimension $\chi$ as the \ac{mps} before the time step.
A simple heuristic method to grow the bond dimension is to replace $\bar{B}^{[n]}$ in step (i) with an isometry $Y_0$ to a larger virtual space, with a dimension $\eta \in [\chi, d\chi]$ which is determined a priori, e.g.~$\eta = \min (\chi_\text{max}, d\chi)$.
In practice, we could take an arbitrary $\eta$-dimensional slice of the left leg pair of $\tilde\theta$.
A controlled method to increase the bond dimension dynamically, based on a desired bound on the truncation error, is given below.
\section{Controlled Bond Expansion}\label{sec:bond_dim_expansion}
\begin{figure}[bt]
    \centering
    \vspace{2ex}
    \includegraphics[width=0.48\textwidth]{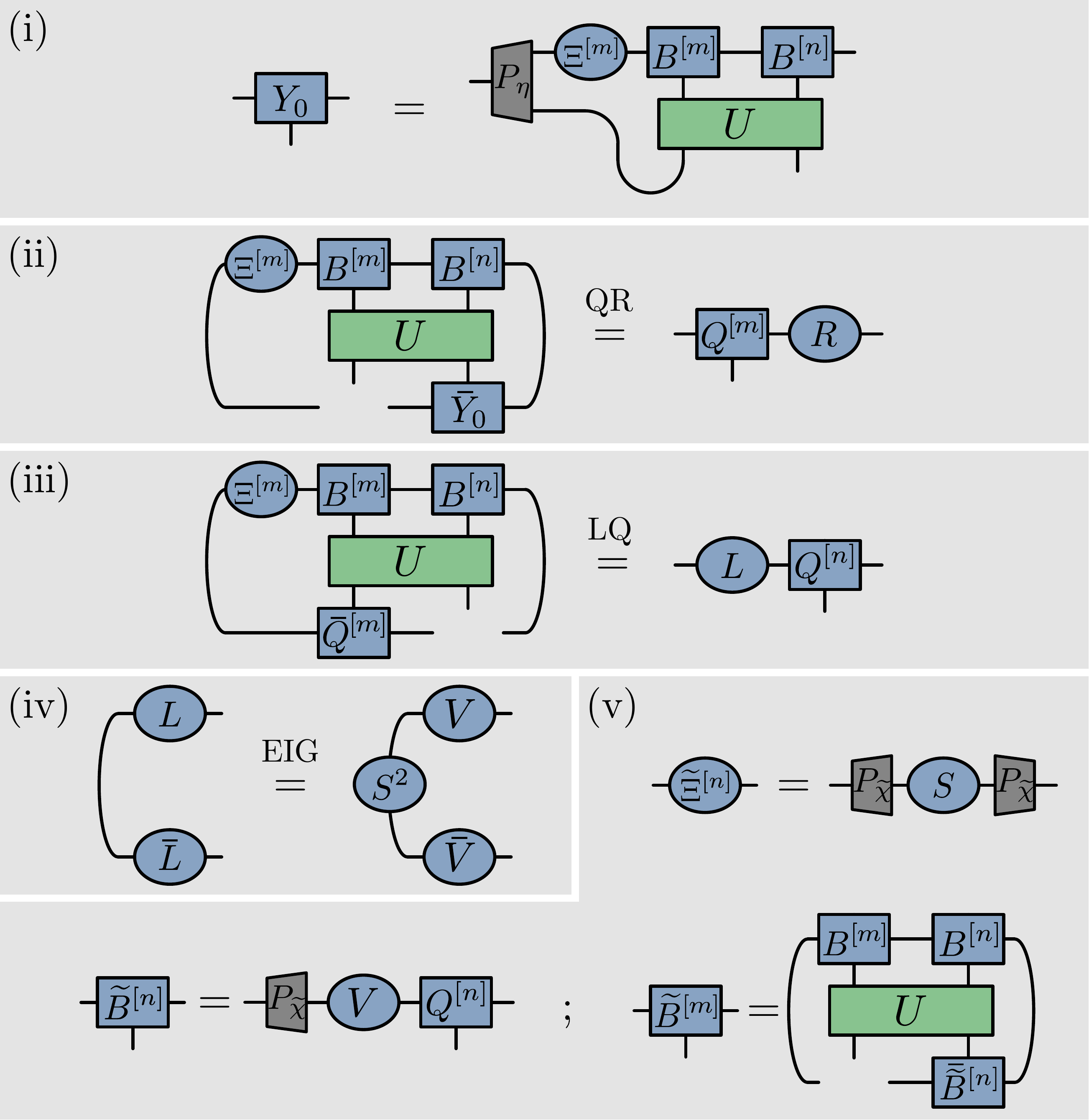}
    \caption{\label{fig:dynamical_bond_dim}
        Algorithm for the QR based time evolution with controlled bond expansion:
        (i) Obtain an initial guess $Y_0$ by projecting / slicing the evolved wavefunction.
        (ii)-(iii) Decomposition of the evolved wave function, similar to section~\ref{sec:algorithm} and Fig.~\ref{fig:algorithm}.
        (iv) Diagonalization of $L^\dagger L$ yields two of the three matrices comprising the \ac{svd} of $L$.
        (v) Truncation of the \ac{svd} to bond dimension $\tilde\chi$, obtaining the updated tensors.
    }
\end{figure}
We discuss now how the \ac{mps} bond dimension can be adjusted dynamically, e.g.~based on the Schmidt values of the state, as can be done in the \ac{svd} based truncation scheme.
This is in analogy to the ideas of controlled bond expansion~\cite{Gleis2022Controlled, Li2022Time, Gleis2022Projector}, which originated in the context of single-site \ac{dmrg}~\cite{White2005Density, Hubig2015Strictly} and improves upon the uncontrolled bond expansion scheme outlined in the previous section.
The algorithm is illustrated schematically in Fig.~\ref{fig:dynamical_bond_dim}.
%
We choose---a priori---a bond dimension $\eta = \chi + \Delta\chi \leq d\chi$ at which we perform the variational QR based decomposition, then truncate to $\tilde\chi \leq \eta$, based on the Schmidt values.
The optimal value of $\Delta\chi$ is model dependent and has to be chosen empirically as the sweet spot in a trade-off between computational cost, which scales as $d^2 \eta \chi^2$, and amount of entanglement which can be represented, for which $\eta$ gives an upper bound.
In practice we find that an increase of $\sim$10 percent at each time step is sufficient for the cases considered.
Next, we require an initial guess for the \ac{mps} tensor on site $n$, which allows to enlarge the dimension of the virtual Hilbert space.
In step (i), we choose an arbitrary $\eta$-dimensional slice on the left leg pair of the time evolved block, i.e.~$(Y_0)^j_{\alpha'\beta} = \sum_{\alpha i} (P_\eta)_{\alpha' (\alpha i)} \tilde{\theta}_{(\alpha i)(\beta j)}$ with the $\eta \times d\chi$ projection matrix $(P_\eta)_{\alpha\beta} = \delta_{\alpha\beta}$.
The following steps (ii) and (iii) involve a QR and a LQ decomposition and are performed exactly as described in the preceding section.
In step (iv), we diagonalize the hermitian matrix $L^\dagger L = V^\dagger S^2 V$, where $S^2$ is a diagonal matrix containing the real, non-negative eigenvalues.
Note, that $S$ are the singular values of $L$ i.e.~the Schmidt values of the state, and we could have computed $S$ and $V$ via an SVD of $L$.
This appears to be significantly slower on the GPU, however.
%
%
%
By discarding the smallest singular values in $S$, along with corresponding rows of $V$, a truncation to a bond dimension of $\tilde\chi$ is achieved, controlled e.g.~by a desired truncation error and/or a threshold below which Schmidt values are neglected.
In step (v), we finally absorb the projected unitary $V$ into $Q^{[n]}$, then update the \ac{mps} tensors as in the previous section, that is $\tilde B^{[n]j}_{\alpha\delta} = \sum_{\beta\gamma} (P_{\tilde\chi})_{\alpha\beta} V_{\beta\gamma} Q^{[n]j}_{\gamma\delta}$, $\tilde\Xi^{[n]}_{\alpha\delta} = (P_{\tilde\chi})_{\alpha\beta} S_{\beta\gamma} (P_{\tilde\chi})_{\gamma\delta}$ and $\tilde{B}^{[m]i}_{\alpha\beta} = \sum_{\gamma\delta i' j' j}U_{ij}^{i'j'}B^{[m]i'}_{\alpha\gamma}B^{[n]j'}_{\gamma\delta}\bar{\tilde{B}}^{[n]j}_{\beta\delta}$, where $P_{\tilde\chi}$ is the projection matrix realizing the truncation, i.e.~keeping the largest $\tilde\chi$ Schmidt values.
If applied to all bonds and if truncation errors are negligible, the bond expansion scheme with its implicit \ac{svd} brings the \ac{mps} to the canonical form, where the $\Xi^{[m]}$ are diagonal matrices containing the Schmidt values.

\par
The controlled bond expansion is crucial when exploiting symmetries via a block structure which these impose~\cite{Singh2010Tensor, Singh2011Tensor}.
The scheme as outlined in section~\ref{sec:algorithm} would need to make a choice about the block structure, in particular the size of the individual blocks, \emph{a priori}, e.g.~using $Y_0 = B^{[n]}$ means choosing them to be the same as before the time step.
The controlled bond expansion, however, allows us to dynamically choose the block dimensions $\chi_i$ optimally, in the sense of minimal truncation error, just like in the \ac{svd} based \ac{tebd} scheme.
Therefore, even when the total bond dimension $\chi=\sum_i \chi_i$ is already saturated, one can expand each block $ \chi_i \rightarrow \chi_i + \Delta \chi_i$
and subsequently truncate back by keeping only at most $\chi$ dominant contributions---this allows to dynamically adjust the size of the individual blocks.
%
%
%
\section{Benchmark}\label{sec:benchmark}
\begin{figure}[t]
    \centering
    \vspace{2ex}
    \includegraphics[width=0.48\textwidth]{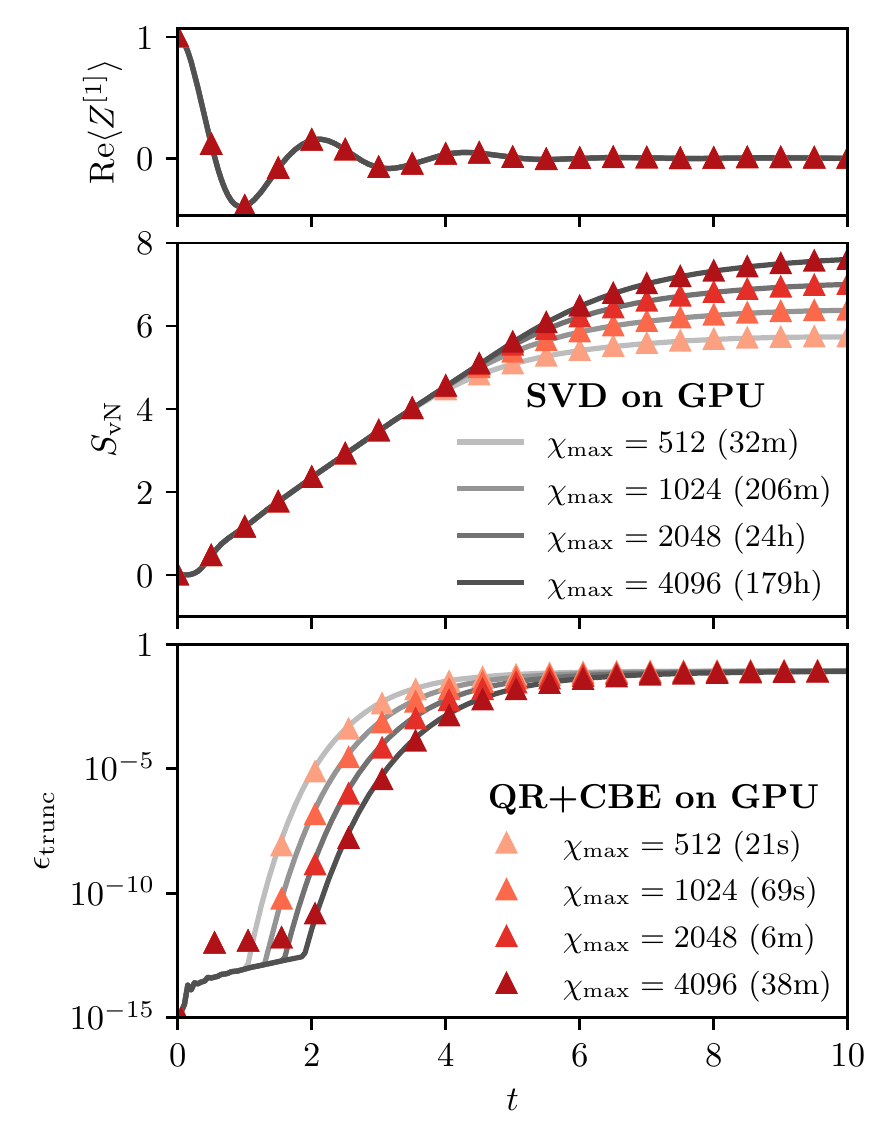}
    \caption{\label{fig:showcase}
        \Ac{tebd} Simulation of a global quench in the $d=5$ quantum clock model from $g=0$ to $g=2$ with a time step of $\delta t = 0.05$ on the A100 GPU.
        We compare local $Z$ expectation values (top), von Neumann entanglement entropy (center) and truncation error (bottom).
        For the QR based scheme, we employ controlled bond expansion with $\Delta\chi = \mathrm{max}(100, 0.1\chi)$ and plot only every tenth datapoint.
        For both schemes, we discard Schmidt values smaller than $10^{-14}$ and keep at most $\chi_\mathrm{max}$ of them.
        Time in the legend denotes the total wall time needed for each simulation, i.e.~to generate the shown data from scratch.
    }
\end{figure}
We choose the $d$-state quantum clock model to benchmark the algorithm.
This model is generically non-integrable ($d>2$) and allows to  highlight the scaling with the physical Hilbert space dimension $d$.
The Hamiltonian reads
\begin{equation}
    H
    = -\sum_{n} \left(Z_n Z_{n+1}^\dagger + \mathrm{h.c.}\right)
    -g \sum_n \left(X_n + \mathrm{h.c.}\right),
\end{equation}
where the clock operators
\begin{equation}
    Z = \begin{pmatrix}
    1 &  & & &
    \\
    & \omega & & &
    \\
    & & \omega^2 & &
    \\
    & & & \ddots &
    \\
    & & & & \omega^{d-1}
    \end{pmatrix}
    ,~
    X = \begin{pmatrix}
    0 & 1 & & &
    \\
     & 0 & 1 & &
    \\
    & & 0 & \ddots &
    \\
    & & & \ddots & 1
    \\
    1 & & & & 0
    \end{pmatrix}
\end{equation}
are $d\times d$ generalization of Pauli matrices and $\omega =\mathrm{e}^{2\pi\mathrm{i}/d}$.
The model has a global $\mathbb{Z}_d$ symmetry generated by $\prod_i X_i$, which we do \emph{not} exploit in the numerical simulations.
For $d \leq 4$, the model has a critical point at $g=1$, while there is an extended critical region around $g=1$ for $d \geq 5$~\cite{Sun2019Phase, Ortiz2012Dualities}.
We start with the $Z=1$ product state and evolve it in time with the $g=2$ Hamiltonian.
\par
We consider four algorithmic variations of the truncation scheme: (i) via \ac{svd}; $\tilde\theta = U S V^\dagger$, (ii) the same decomposition but numerically evaluated by diagonalizing $\tilde\theta^\dagger \tilde\theta = V S^2 V^\dagger$ (note that $U$ is not actually required), which we dub EIG, (iii) the simple QR based scheme we have introduced in section~\ref{sec:algorithm}, and (iv) the QR scheme with controlled bond expansion (QR+CBE) as described in section~\ref{sec:bond_dim_expansion}.
We run the benchmark on a NVIDIA A100 GPU (80GB RAM) with CUDA version 11.7, as well as an AMD EPYC 7763 CPU with 64 physical cores and MKL version 2019.0.5.
The two units have similar power consumption; $300\mathrm{W}$ and $280\mathrm{W}$ thermal design power respectively.
All simulations are performed in double precision (i.e., complex128).
\par
In Fig.~\ref{fig:showcase}, we perform full \ac{tebd} simulations of the quench protocol for a $d=5$ clock model.
We run the simulation beyond times where the approximation of the evolved state as an \ac{mps} of the given bond-dimension breaks down, as quantified by a large truncation error.
In the time regime of acceptable error $\epsilon_\text{trunc} \lesssim 10^{-5}$, that is until $t = \lesssim 2$ depending on bond dimension, we observe excellent agreement between the different \ac{tebd} schemes in all extracted quantities up to relative deviations of $10^{-11} \sim 10^{-12}$.
For the QR based scheme, we do not have access to all singular values of $\tilde\theta$, from which the truncation error is extracted in \ac{svd} based \ac{tebd}.
We instead explicitly compute the distance between the evolved wave function $\tilde\theta$ and its low rank approximation.
\par
\begin{figure}[hbt]
    \centering
    \vspace{2ex}
    \includegraphics[width=0.48\textwidth]{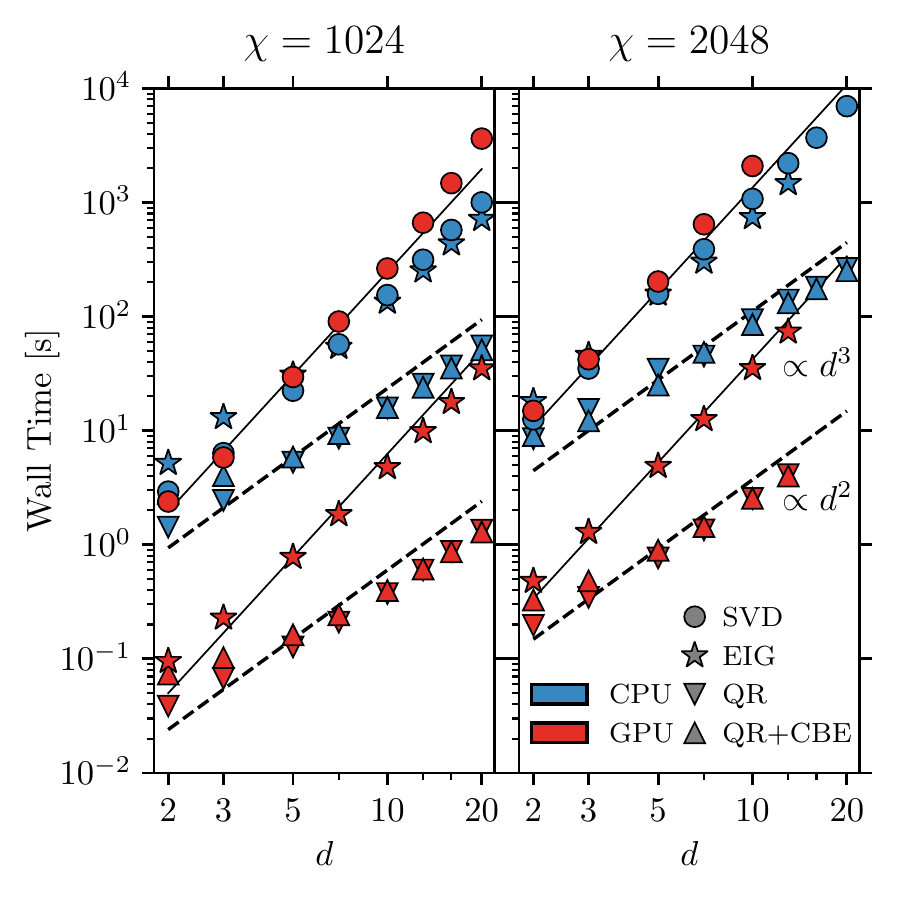}
    \caption{\label{fig:benchmark}
        Timing benchmark for the application of a single gate to an \ac{mps} for different hardware (marker colors) and truncation schemes (marker shapes).
        We give the average wall time needed to compute the updated tensors $\tilde{B}^{[m]}$, $\tilde{\Xi}^{[n]}$, $\tilde{B}^{[n]}$ from the old tensors ${\Xi}^{[m]}$, ${B}^{[m]}$, ${B}^{[n]}$ and $U$.
        For the QR based algorithm with (without) controlled bond expansion (CBE), these are the steps illustrated in Fig.~\ref{fig:dynamical_bond_dim}~(Fig.~\ref{fig:algorithm}) and $\Xi$ are diagonal (non-diagonal) matrices.
        The task consists of first contracting $\tilde\theta$, performing an SVD of $\tilde\theta$ (diagonalizing $\tilde\theta^\dagger\tilde\theta$) and finally contracting $\tilde B^{[m]}$ in the case of SVD (EIG) based truncation.
        For CBE, we choose $\Delta\chi = 0.1 \chi$, the same expansion rate as for Fig.~\ref{fig:showcase}.
        The initial MPS has a bond dimension $\chi$ and the evolved state is truncated to $\tilde\chi = \chi$.
        Solid (dashed) lines are powerlaws with the expected cubic (quadratic) scaling with the physical dimension $d$.
        The missing datapoints for large $d$ in the right panel were not possible to obtain due to memory limitations.
    }
\end{figure}

In Fig.~\ref{fig:benchmark}, we benchmark runtimes for the core algorithmic step of contracting and subsequently decomposing the evolved wavefunction $\tilde\theta$ for all combinations of truncation scheme and hardware, as well as a range of Hilbert space dimensions $d$.
We clearly observe the improved scaling of the QR based algorithm which is quadratic in $d$ instead of cubic, as well as speed-up of one to two orders of magnitude from hardware acceleration for EIG and QR based algorithms.
Notably, the \ac{svd} truncation is instead slower on the GPU.
For example, the QR based truncation scheme on the GPU with $\chi = 1024$, $d=20$ reaches a speed-up factor of $2700$ compared to the SVD based scheme on the same GPU and $750$ compared to SVD on CPU.
%
%
%
\section{Conclusion}\label{sec:conclusion}
We proposed and benchmarked a modified time evolution block decimation (TEBD) algorithm that
uses a truncation scheme based on the QR decomposition instead of the singular value decomposition (SVD).
We demonstrated that the QR based truncation scheme allows simulation of the time evolution of MPS to the same degree of accuracy, but compared to the \ac{svd} based \ac{tebd} scheme drastically decreases runtime and power consumption needed to obtain the same results, especially when run on GPU hardware.
\par
We expect that with small changes, the algorithm can be used to accelerate \ac{mps} truncation in a broader class of algorithmic settings, e.g.~to apply long range gates arising in the effective 1D description of two dimensional models, time evolution based on applying MPOs~\cite{Zaletel2015Time}, or DMRG~\cite{White1992}.
An application to the simulation of quantum circuits would need to be investigated in further detail, since unlike for the Trotterized time evolution with small time steps, the unitary gates in a generic quantum circuit need not be close to unity.
Hardware acceleration on the heavily specialized tensor processing units (TPUs)~\cite{Ganahl2022, lewis2022large} may yield an even greater performance increase and make larger bond dimensions accessible via large memory and distributed linear algebra, allowing the simulation to represent more entanglement.
For the simulation of finite systems, parallel gate application can provide further performance increase, as demonstrated in Ref.~\cite{Secular2020} for the time dependent variational principle (TDVP).
%
%
%
%
%
\begin{acknowledgements}
This research was financially supported by the European Research Council (ERC) under the European Union’s Horizon 2020 research and innovation program under grant agreement No. 771537. F.P. acknowledges the support of the Deutsche Forschungsgemeinschaft (DFG, German Research Foundation) under Germany’s Excellence Strategy EXC-2111-390814868. F.P.’s research is part of the Munich Quantum Valley, which is supported by the Bavarian state government with funds from the Hightech Agenda Bayern Plus.

\textbf{Data and materials availability:} – Data analysis and simulation codes are available on Zenodo~\cite{zenodo}.
\end{acknowledgements}

%
%

\bibliography{main}

\end{document}